\documentclass[12pt]{iopart} 
\usepackage{epsfig,multicol,iopams,graphicx}

\def \pre{Preprint}

\begin{document}

\title{Area spectra of the rotating BTZ black hole from quasinormal modes}

\author{Yongjoon Kwon and Soonkeon Nam}

\address{ Department of Physics and Research Institute of Basic Sciences, \\
 Kyung Hee University, Seoul 130-701, Korea}

\eads{\mailto{emwave@khu.ac.kr}, \mailto{nam@khu.ac.kr}}



\begin{abstract}
Following Bekenstein's suggestion that the horizon area of a black hole should be quantized, the discrete spectrum of the horizon area has been investigated in various ways. By considering the quasinormal mode of a black hole, we obtain the transition frequency of the black hole, analogous to the case of a hydrogen atom, in the semiclassical limit.  According to Bohr's correspondence principle, this transition frequency at large quantum number is equal to classical oscillation frequency. For the corresponding classical system of periodic motion with this oscillation frequency, an {\it {action variable}} is identified and quantized via Bohr-Sommerfeld quantization, from which the quantized spectrum of the horizon area is obtained. This method can be applied for black holes with discrete quasinormal modes. As an example, we apply the method for the both non-rotating and rotating BTZ black holes and obtain that the spectrum of the horizon area  is equally spaced and independent of the cosmological constant for both cases.
\end{abstract}

%


\maketitle


%
\section{Introduction}
It is believed that the quantum behavior of black holes could play a significant role as a testing ground in understanding the quantum theory of gravity. 
A few decades ago Bekenstein \cite{be} showed that the black hole horizon area is an adiabatic invariant, so that it should be quantized  as follows: By Ehrenfest principle  \cite{ehren}  that only an adiabatically invariant quantities are quantizable
\begin{equation}
A_n = f(n)   ~~,~~ n=0,1,2..~ .
\end{equation}
By considering the minimum change of the horizon area, $\triangle A_{\rm {min}}$, by an absorption of a test particle into the black hole, Bekenstein argued that $\triangle A_{\rm {min}}$ can only be proportional to $\hbar$, so that the area spectrum should be linearly quantized as follows \cite{be}: In the  $c=G=1$ units 
\begin{equation}
A_n =\gamma \, n \,\hbar =\gamma \, n\, { \ell_p ^2} ~~,~~ n=0,1,2..~ ,
\end{equation}
where $\gamma$ is an undetermined dimensionless constant and $\ell_p= \sqrt{{ {\hbar \,G} \over {c^3} }}$ is the Planck length.
This  equally spaced area spectrum can help us understand why the entropy of a black hole is proportional to the horizon area in intuitive way. It can be considered that the horizon is formed by patches of equal area $\gamma \,{\ell_p ^2}$ and the patches get added one at a time \cite{be}. Since this proposal, there have been a lot of works on the area spectrum of black holes in various ways \cite{old, hod, kun, recent, magi, setvag,  vage, med, setare, wei, wei2}. 
Among these, methods using the quasinormal mode of a black hole, which is the characteristic sound of the black hole, have been investigated \cite{hod, kun, recent, magi, setvag,  vage, med, setare, wei, wei2}. 
A significant interpretation of quasinormal mode in quantizing the horizon area of a black hole was done by Hod \cite{hod}, who suggested that the dimensionless constant $\gamma$ can be determined by using the quasinormal mode of a black hole \cite{hod}. 
The real part of the asymptotic quasinormal mode was regarded as a transition frequency in the semiclassical limit and the quantum emission of the transition frequency results in a change of the black hole mass, $\triangle M = \hbar \,\omega_R$.  
For  the Schwarzschild black hole, the asymptotic quasinormal mode is given by \cite{hod, nol}
\begin{equation}
\omega _k = {{{\rm ln}\,3} \over {8\,\pi\,M}} - {i \over {4\,M}} \, \left( k+{1 \over 2} \,\right) +O(k^{-1/2} )  ~. %
\end{equation}
Using the relations $A=16\,\pi \,M^2$  and $\triangle M=\hbar\, \omega_R= {{ \hbar\,{ \rm ln\,3} } \over {8\,\pi\,M}} $,  the dimensionless constant $\gamma$ is determined as
\begin{equation}
\gamma = 4\, {\rm ln\, 3 } ~.
\end{equation}

Later, Kunstatter \cite{kun}  quantized the horizon area of the black hole by using an adiabatic invariant $I$.
It was noticed that in a system with energy $E$ and vibrational frequency $\omega(E)$,  the quantity $I=  \int {dE / {\omega(E)} } $ is an adiabatic invariant \cite{kun}. The adiabatic invariant $I$ was quantized via Bohr-Sommerfeld quantization. To obtain the quantization of the horizon area for the Schwarzschild black hole, the real part of the quasinormal frequency was used as the vibrational frequency. Therefore with the energy of the black hole given by the ADM mass $M$, $E=M$, the quantization of the adiabatic invariant is given by  \cite{kun} 
\begin{equation} \label{kuf}
I = \int {dE \over {\omega _R} } = \int {dM \over {\omega _R} } = n \, \hbar ~.
\end{equation}
Since the real part of the quasinormal mode  for the Schwarzschild black hole is given by $\omega _R= { { {\rm ln} \,3} \over {8\,\pi\,M}}$, the above formula gives mass quantization 
\begin{equation}
M^2 = { {{\rm ln}\,3} \over{4\,\pi\,} } \,n \, \hbar ~,
\end{equation}
which leads to the quantization of the horizon area in the semiclassical limit.
Therefore the area spectrum of the Schwarzschild black hole is equally spaced and agrees with Hod's result \cite{hod}, reproducing $\gamma =  4\,{\rm ln}\,3$.

More recently, Maggiore \cite{magi} proposed that we have to describe the quasinormal mode of a {\it {perturbed}} black hole as a {\it damped} harmonic oscillator because the quasinormal mode has an imaginary part. It was proposed that we cannot neglect the imaginary part of the quasinormal frequency, and it was noticed that the proper physical frequency of the harmonic oscillator with a damping term should be given by
\begin{equation}
\omega _0 ={\sqrt{ {\omega _R}^2+{\omega _I}^2 } } ~.
\end{equation}
For the highly excited modes of most black holes, the imaginary part is much larger than the real part, $\omega _{I} \gg \omega _{R}$. Therefore we have  $\omega _0 \sim \omega _I  $ rather than $ \omega _0 \sim \omega _R$ as has been used before  \cite{magi}. 
Furthermore, it was proposed that the characteristic classical frequency $\omega _c$ should be identified with the transition frequency between the quantum levels $(\omega _{0})_k $ of the black hole in the semiclassical limit \cite{magi}.  That is, for highly damped modes ($\omega _I \gg \omega _R$) we have
\begin{eqnarray}
\omega _{c} &=& (\omega _ {0})_{k} - (\omega _ {0})_{k-1}   \nonumber  \\
&\simeq & ( | \omega _ {I} | )_{k} - ( | \omega _ {I} | )_{k-1}    ~~,~~   (k \in N,~~k \gg 1) ~.
\end{eqnarray}
The transition gives rise to the emission of a quantum, which results in change of the mass of a black hole.
With this identification, the area spectrum of the Schwarzschild black hole becomes  $A=8\,\pi\,n\,\hbar$, using the relation $\triangle M = \hbar\,\omega _c$. This spectrum is again equally spaced, but the spacing is different. 
Some authors \cite{med, wei} obtained the same area spectrum for the Schwarzschild black hole by using the formula (\ref{kuf}) with the newly identified frequency $\omega _c$ instead of $\omega _R$.
%

We can also consider the quantization of the horizon area for a rotating black hole.  
There were some attempts to quantize it  using the adiabatic invariant \cite{setvag, vage, med}.  For example, a modified adiabatic invariant following Kunstatter's proposal \cite{kun} was considered and the real part of the quasinormal mode was used as the vibrational frequency \cite{setvag}. So, the modified adiabatic invariant ${\mathcal I}$ via Bohr-Sommerfeld quantization is given by
\begin{equation} \label{setv}
{\mathcal I} = \int {{dM -\Omega \, dJ}  \over {\omega _R} }= n\,\hbar ~,
\end{equation}
where $\Omega$ is the angular velocity at horizon and $J$ is the angular momentum of the black hole.
This adiabatic invariant for the Kerr black hole does not give equally spaced area spectrum \cite{setvag}. Later,  the Maggiore's transition frequency $\omega _c$, instead of $\omega _R$, was applied as the vibrational frequency to the adiabatic invariant (\ref{setv}) for the Kerr black hole \cite{vage, med}. Here again, the horizon area is not equally spaced. Only for the large mass compared to the angular momentum, it is shown that $  A \simeq 8 \,\pi \,n\,\hbar $ approximately  \cite{med}. The result of the non-equally spaced area spectrum is not consistent with Bekenstein's proposal \cite{be}. 

In this work we will find that the area spectrum of a rotating black hole is also equally spaced by identifying the corresponding classical system and using the Bohr-Sommerfeld quantization.
In order to quantize a classical system via Bohr-Sommerfeld quantization in the semiclassical limit we would like to point out that an action variable rather than an adiabatic invariant should be quantized. Of course, in certain cases these two can be identical, but they differ in general. 
So, we will find an action variable of a classical mechanical system corresponding to a quantum black hole, based on Bohr's correspondence principle.

\section{Black hole as an oscillator and its semiclassical quantization}
In previous works \cite{setvag, vage, med}, from the Bekenstein's suggestion that the horizon area is an adiabatic invariant, the modified form of the adiabatic invariant ${\mathcal I}$  was obtained as follows:
\begin{equation}
{\mathcal I} = \int \,dA \sim \,\int \,{{ {T_H}\,dS} \over {T_H} } \sim \,\int \,{{dM-\Omega \,dJ} \over \omega} ~,
\end{equation}
where Bekenstein-Hawking area law \cite{area}, the first law of black hole thermodynamics \cite{bar} and $\omega \sim {T_H}$ are used.
To quantize the modified adiabatic invariant, Bohr-Sommerfeld quantization was applied as follows:
\begin{equation}
{\mathcal I} =\int \,{{dM-\Omega \,dJ} \over \omega}= n \,\hbar ~.
\end{equation}
Here, we would like to remind ourselves of why the adiabatic invariant was quantized in the first place.
According to Ehrenfest principle \cite{ehren},  any classical adiabatic invariant corresponds to a quantum entity with discrete spectrum. However it does not mean that the adiabatic invariant ${\mathcal I}$ is always equally spaced or quantized as ${\mathcal I}=n\,\hbar$. Only when an adiabatic invariant $\mathfrak I$ is an action variable of a classical system,  Bohr-Sommerfeld quantization is applied as follows:
\begin{equation} \label{bs}
{\mathfrak I}= {1 \over {2\,\pi}}\,\oint p\,dq = n\,\hbar  ~~,~~ n=0,1,2,..\,.
\end{equation}
An action variable of the classical system is an adiabatic invariant, but not every adiabatic invariant is an action variable. For example, if $\mathcal K$ is an action variable which is equally spaced via Bohr-Sommerfeld quantization, $\mathcal K^2$ is not equally spaced even though it is an adiabatic invariant. 
%
%
To see this point, we would like to reconsider the example of a rotating black hole. For this black hole, the adiabatic invariant is ${\mathcal I} =\int \,{{dM-\Omega \,dJ} \over \omega}$. However we should not quantize this quantity via Bohr-Sommerfeld quantization. Rather we have to  identify a classical mechanical system whose action variable can be easily calculated.
We claim that the action variable should be ${\mathfrak I}= {\int { {dM} \over {\omega _c}} }$, where $M$ is the ADM mass of the black hole and $\omega _c$ is the transition frequency.

Our argument is as follows: We will consider the analogy to the hydrogen atom and obtain the transition frequency $\omega _c$ of the quantum black hole from quasinormal modes in the semiclassical limit. 
A hydrogen atom emits a photon in the form of radiation  when the transition between the states occurs. Likewise, in analogy with the atomic transition, a quantum black hole system emits a quantum in the form of radiation when the transition between the states occurs. 
Bohr's correspondence principle says that the transition frequency at large quantum number is equal to classical oscillation frequency.  
Therefore based on this correspondence principle {\it{the quantum black hole with the transition frequency  $\omega _{c}$ can be regarded as a classical system of periodic motion with oscillation frequency $\omega _c$ in the semiclassical limit}}.

For a classical system of periodic motion
we can consider an action-angle variable in a system of one degree of freedom with oscillation frequency $\omega _c$ of periodic motion \cite{gol}.
Since we consider a conservative classical system, the Hamiltonian $H$ can be written as $H(p,q)=E\equiv \alpha$, where $\alpha$ is a constant.
By a canonical transformation we introduce a new variable $\mathfrak I$ as a generalized momentum, which is an action variable  defined as 
\begin{equation} \label{ham}
{\mathfrak I}= {1 \over {2\,\pi}}\,\oint p\,dq  ~.
\end{equation}
Since $p=p\,(\alpha, q)$, this action variable $\mathfrak I$ is a function of $\alpha$ only;
\begin{equation} 
H =H({\mathfrak I}) \equiv \alpha   ~.
\end{equation}
With the angle variable $\theta$, which is the generalized coordinate  conjugate to ${\mathfrak I}$, the Hamilton equations of motion read 
\begin{equation} \label{ham3}
{\dot \theta} = { {\partial H({\mathfrak I}) } \over {\partial {\mathfrak I} }} = \beta ({\mathfrak I}) ~~,~~ \dot {\mathfrak I} = - { {\partial H} \over {\partial \theta}}=0 ~,
\end{equation}
where $\beta$ is a constant function of $ {\mathfrak I} $.
Therefore the action variable ${\mathfrak I}$ is a constant of motion and the angle variable $\theta$ is given by
\begin{equation}  \label{ham5}
\theta = \beta \,t + {\rm constant} ~.
\end{equation}
One can show that the change of $\theta$ for one cycle of the periodic motion is $2\,\pi$ \cite{gol}. Therefore from (\ref{ham5}) we obtain 
\begin{equation} 
\triangle \, \theta = 2\,\pi = \beta\,\tau  ~,
\end{equation}
where $\tau$ is the period of one cycle of the periodic motion. 
Since from this relation  the constant $\beta$ is given  by 
\begin{equation}
\beta ={ {2\,\pi} \over {\tau}} ~,
\end{equation}
the $\beta$ is the oscillation (angular) frequency $\omega _c$ of the periodic motion;
\begin{equation} \label{ham4}
\beta = \omega _c~.
\end{equation}
Hence from (\ref{ham3}) and (\ref{ham4}) we obtain the following relation:
\begin{equation}
dH =\omega _c \,d{\mathfrak I} = dE ~.
\end{equation}
Integrating the both sides of $d{\mathfrak I} = dE/{ \omega _c}$,  the action variable ${\mathfrak I}$ of the periodic system with the frequency $\omega _c $  is written as
\begin{eqnarray}
{\mathfrak I} =\, \int { dE \over {\omega _c }}  ~. 
\end{eqnarray}
This action variable can be quantized via Bohr-Sommerfeld quantization as in (\ref{bs}). Because the change of the energy $E$ of the system is the change of the mass $ M$ (ADM mass) of the black hole, we finally obtain, in the semiclassical limit, 
\begin{equation} \label{btz1}
{\mathfrak I} =\, \int { dM \over {\omega _c}}  = n\, \hbar ~.
\end{equation}
This is exactly the same formula as what was used for the Schwarzschild black hole in \cite{med, wei}.
Since we can also view other black holes with discrete quasinormal mode as a system with transition frequency $\omega _c$ in the semiclassical limit, our claim is that we can also apply this formula (\ref{btz1}) to other black holes. 
Based on Bohr's correspondence principle,
we can regard these black holes as oscillators with the frequency $\omega _c$.
Once we obtain the action variables of these oscillators, we can quantize them via Bohr-Sommerfeld quantization.
As an example, we will apply the above argument for the rotating BTZ black hole, which has the exactly solved quasinormal modes.
First, we will consider the non-rotating BTZ black hole with one horizon and show that the horizon area spectrum is equally spaced and independent of the cosmological constant.  Next, we will consider the rotating BTZ black hole with two horizons and find that the inner horizon area as well as the outer horizon area is equally spaced and independent of the cosmological constant. We will find that the entropy spectrum is also equally spaced. 
From now on, we will use the gravitational BTZ units in which $ c=8\,G=1$.

\section{Area spectrum of the non-rotating BTZ black hole}   
First, we would like to apply the method mentioned in previous section to the non-rotating BTZ black hole whose metric is given as follows \cite{BTZ}:
\begin{equation} \label{met}
ds^2=-\left(-M+{ r^2 \over l^2} \right)\,dt^2+ { {dr^2} \over {\left(-M+{ r^2 \over l^2} \right)} } +r^2\,d\phi ^2 ~,
\end{equation}
where the cosmological constant is given by $\Lambda=-{1 \over l^2}$.
The  horizon is located at $r_H = l\, \sqrt{M}$.
In a previous work the real part of the quasinormal frequency was used and non-equally spaced area spectrum was obtained \cite{setare}.
In this paper, however, we are using the transition frequency $ \omega _c$.

The quasinormal mode for the non-rotating BTZ black hole is given by  \cite{cardo, bir}
\begin{equation} \label{qn}
\omega =\pm {\frac {m}{l}}-2\,{\frac {i\sqrt {M} \left( k+1 \right) }{l}}  ~~,~~ (m \in Z \,,~ k =0,1,2,..)~,
\end{equation}
where $m$ is the angular quantum number and $k$ is the overtone quantum number of the quasinormal mode which comes from the boundary condition in the radial direction.

At large $k$ for a fixed $\vert m \vert$, in particular for $k \gg \vert m \vert$, the proper physical frequency $\omega _0$
 is dominated by the imaginary part of the quasinormal frequency; 
\begin{equation}
\omega _0 \simeq (\vert \omega _I \vert)_k = { {2 \, \sqrt{M} \, (k+1)} \over l} ~.
\end{equation}
The transition frequency $\omega _c$   between two highly excited neighboring states becomes
\begin{equation}
\omega _c \simeq ( \vert \omega _I \vert )_{k} - ( \vert \omega _I \vert )_{k-1} = {{ 2\, \sqrt{M} } \over {l}} ~.
\end{equation}
Applying this to (\ref{btz1}), we get
\begin{equation}
{\mathfrak I} = \int  { dM \over {\omega _c} } = {l \over 2} \int { {dM} \over {\sqrt{M}} } = l\,\sqrt{M} ~.
\end{equation}
Thus the action variable is just the horizon radius $r_H$.
Therefore the horizon radius is quantized via Bohr-Sommerfeld quantization;
\begin{equation}
r_H = l \,\sqrt{M}  = n \,\hbar  ~.
\end{equation}
Finally, the horizon area is quantized and equally spaced as follows:
\begin{equation}
A =2\,\pi\,r_H = 2 \,\pi \,n \,\hbar  ~.
\end{equation}
The quantization of the entropy is also equally spaced;
\begin{equation} \label{bb1}
\triangle S = 4 \,\pi  ~,
\end{equation}
because the entropy is given by Bekenstein-Hawking area law \cite{area}; 
\begin{equation} \label{alaw}
S={{4\,\pi \,r_{+}} \over \hbar}={ {2\,A} \over  \hbar} ~~
\end{equation}
in the units where $c=8\,G=1$.
The area and entropy spectra are independent of the cosmological constant $\Lambda =-1/l^2$, consistent with the results for other AdS black holes \cite{wei2}. 
Apparently our result is different from the previous result in \cite{wei}, where the quasinormal mode taken $l=1$  was used and the area  spectrum was dependent of the cosmological constant.  One actually has to use the $l$-dependent quasinormal mode of (\ref{qn}) for the metric (\ref{met}), and the puzzle that the area spectra of the non-rotating BTZ black hole obtained  in \cite{wei}  looks different from other AdS black holes whose area spectra are independent of the cosmological constant \cite{wei2} is actually resolved. 
Furthermore one has to use $c=8\,G=1$ and the correct relation between area and entropy should be given by (\ref{alaw}).

\section{Area spectra of the rotating BTZ black hole}
Now, we would like to consider the rotating BTZ black hole whose area spectrum was not calculated from quasinormal modes before.
The metric of the rotating BTZ black hole  is given by \cite{BTZ}
\begin{eqnarray}
ds^2 =& -&\left(-M+{ r^2 \over l^2} +{ J^2 \over {4\,r^2} } \right)dt^2+ { {dr^2} \over {\left(-M+{ r^2 \over l^2} +{ J^2 \over {4\,r^2} }\right)} } \\
&+&\, r^2 \, \left( d\phi -{J \over {2\,r^2} }\,dt\right) ^2 ~,
\end{eqnarray}
where the cosmological constant is given by $\Lambda=-{1 \over l^2}$.
The mass $M$ and angular momentum $J$ of the black hole can be expressed in terms of the outer and inner horizons, $r_\pm$, as follows:
\begin{equation}
M= { {  {r_+ ^2}+ {r_- ^2} } \over l^2} ~~,~~ J= { {  2\,{r_+}\, {r_-} } \over l} ~.
\end{equation}

The two families of the quasinormal modes of the rotating BTZ black hole for a massive scalar field 
are given by \cite{bir}
\begin{eqnarray}
\omega _{R} &=& -{\frac {m}{l}}- i {\frac { (r_{+}+r_{-} ) \left( 2 k+1+{\sqrt {1+\mu \,} } 
 \right) }{{l}^{2}}} ~,\\
\omega _{L} &=& {\frac {m}{l}}- i {\frac { (r_{+}-r_{-} ) \left( 2 k+1+{\sqrt {1+\mu \,} } 
 \right) }{{l}^{2}}}   ~,
\end{eqnarray}
where $m \in Z \, $ and $ \, k=0,1,2,..$.
The $m$ and $ k$ are the angular quantum number and the overtone quantum number respectively, and $\mu$ is the mass parameter,  $\mu \equiv u^2\,l^2 / {\hbar ^2}$, where $u$ is the mass of the scalar field.
At large $k$ for a fixed $\vert m \vert$, in particular for $k \gg \vert m \vert$, the two families of quasinormal modes gives two possible transition frequencies, $\omega _{Rc}$ and $ \omega _{Lc}$.
We find the two transition frequencies corresponding to each quasinormal mode;
\begin{eqnarray}
\omega _{R c} &=&  {{2\, (r_{+}+r_{-} ) } \over {l^2}} = { {2 \sqrt{M+{J / l} \, } } \over l} ~~,\\
\omega _{L c} &=&  {{2 \,(r_{+}-r_{-} ) } \over {l^2}} = { {2 \sqrt{M-{J / l} \, } } \over l}  ~.
\end{eqnarray}
We consider the two action variables corresponding to each possible transition frequency. 
From the formula (\ref{btz1}),
we obtain the two quantization conditions;
\begin{equation} \label{t2}
{\mathfrak I_{R}} = \int  { dM \over {\omega _{Rc} } }  = l \sqrt{ M+J/l \, } = n_R \, \hbar ~,
\end{equation}
\begin{equation} \label{t3}
{\mathfrak I_{L}} = \int  { dM \over {\omega _{Lc} } }  = l \sqrt{ M-J/l \,} =n_L \, \hbar  ~,
\end{equation}
where $n_R, \, n_L= 0,1,2,..$ . Notice that we have $n_R \ge n_L$. 

Because the total horizon area is given by 
\begin{equation}
A_{tot} \equiv A_{out}+A_ {in} = 2 \,\pi \,l \, \sqrt{ M+J/l \,}  ~,
\end{equation}
we find that the total horizon area is quantized  and equally spaced;
\begin{equation} \label{b1}
A_{tot}=2 \,\pi \,n_R \, \hbar  ~~,\quad ~~n_R =0,1,2,..\,.
\end{equation}
On the other hand, the difference of the two horizon areas is given by 
\begin{equation} 
A_{sub} \equiv A_{out}-A_ {in} = 2\, \pi \,l \, \sqrt{ M-J/l \,}  ~.
\end{equation}
Therefore the area difference is quantized and also equally spaced;
\begin{equation}  \label{b2}
A_{sub}=2 \,\pi \,n_L\, \hbar   ~~,\quad ~~n_L=0,1,2,..\,.
\end{equation}
From (\ref{b1}) and (\ref{b2}), we find the quantizations of the outer and inner horizon areas;
\begin{equation} 
A_{out}=\,\pi \, (n_R + n_L ) \,\hbar ~~,  ~~A_{in}= \,\pi \, (n_R - n_L) \,\hbar  ~,
\end{equation}
where $ n_R \ge n_L\,$.
Therefore we find that both the outer and inner horizon areas are equally spaced with the same spacing; 
\begin{equation} 
\triangle A_{out}=\,\pi \, \hbar ~~,  ~~ \triangle A_{in}= \,\pi  \, \hbar  ~.
\end{equation}
By Bekenstein-Hawking area law, the entropy spectrum  is given by
 \begin{equation} \label{t7}
S= { {4 \, \pi \, r_{+} } \over \hbar} = { {A_{tot}+A_{sub} } \over \hbar}= 2 \,\pi  \,(n_R + n_L)  ~.
\end{equation}
Hence it also has equal spacing,  consistent with Bekenstein's proposal;
\begin{equation} \label{b_3}
\triangle S= 2\,\pi ~.
\end{equation}
Notice that the spacing of the entropy spectrum is half of the non-rotating BTZ black hole case and same as for the Schwarzschild black hole \cite{med, wei}.

Let us now consider the case of  when the angular momentum $J$ goes to zero.  
Because the inner horizon area goes to zero, the total horizon area is given by
\begin{equation}
A_{tot} \simeq A_{out} \equiv A ~.
\end{equation}
Furthermore from (\ref{t2}) and (\ref{t3}) $n_L$ goes to $n_R$.
This means that $A_{out} =\pi \, (n_R + n_L)\,\hbar  \simeq 2\,\pi \, n_R \,\hbar $ . 
Therefore the spectrum of the single horizon area is equally spaced;
\begin{equation}
\triangle A= 2 \,\pi \, \hbar  ~.
\end{equation}
The spacing of the entropy spectrum is
\begin{equation}
\triangle S= 4 \,\pi   ~,
\end{equation}
since $n_L \simeq n_R$.
This is consistent with the result (\ref{bb1}) of the non-rotating BTZ black hole.

\section{Conclusion}

We calculated the area spectra of  the both non-rotating and rotating BTZ black holes by quantizing the action variables of the corresponding classical mechanical systems via Bohr-Sommerfeld quantization. 
For this, we considered black hole systems with discrete quasinormal modes as follows:
From the discrete quasinormal mode of a black hole we obtained the transition frequency $\omega _c$ of the black hole in the semiclassical limit. In analogy to the atomic transition, the black hole with the transition frequency was considered as a `hydrogen atom-like' quantum black hole, where a transition between neighboring states gives rise to the emission of a quantum of that frequency. By Bohr's correspondence principle which says that the transition frequency at large quantum number is equal to the classical oscillation frequency, the quantum black hole with the transition frequency $\omega _c$ was regarded as a classical system of periodic motion with the oscillation frequency $\omega _c$.
Using the action variable ${\mathfrak I}$ of the system of periodic motion, we quantized it via  Bohr-Sommerfeld quantization in the semiclassical limit as follows:  ${\mathfrak I} = {1 \over {2\,\pi}}\,\oint p\,dq = \int {{dE } \over {\omega _c}}= n\,\hbar$.  
Since the change of the energy $E$ of a black hole is the change of the ADM mass $M$,  we obtained the formula, $ \int {{dM } \over {\omega _c}}= n\,\hbar$, which gives the quantization condition of the black hole. 
This formula can be applied regardless of whether a black hole is rotating or not.  
The transition we have considered was between highly excited levels for fixed orbital quantum number, so it does not  change  the angular momentum $J$ of a black hole. 
With this formula  we found that the both non-rotating and rotating BTZ black holes have equally spaced area spectra, which are consistent with Bekenstein's proposal that the horizon area should be equally spaced \cite{be}. 

We obtained the following results:
For the non-rotating BTZ black hole there is one action variable corresponding to  one transition frequency $\omega _c$. 
From Bohr-Sommerfeld quantization of this action variable it was found that the area and entropy spectra are equally spaced and independent of the cosmological constant $\Lambda=-1/l^2$, different from the previous results for the non-rotating BTZ black hole in \cite{wei}. The previous results  were not consistent with  the results of other AdS black holes in \cite{wei2}, which have the area spectrum independent of the cosmological constant. But our results are consistent with the results.  
For the rotating BTZ black hole  two action variables are obtained, because there are two possible transition frequencies, $\omega_{Rc}$ and $\omega_{Lc}$. Via  Bohr-Sommerfeld quantization, each  action variable results in the quantization of the total horizon area and the quantization of the difference between  two horizon areas. From those quantization conditions we obtained that {\it {the outer and inner horizon areas are quantized and the both area spectra are equally spaced}}. 
By Bekenstein-Hawking area law, the entropy spectrum is also equally spaced. 

In this work, we obtained the area spectra of non-extremal BTZ black holes through semiclassical quantization. Since the quasinormal mode is absent in extremal BTZ black hole, we cannot find the area spectrum of extremal BTZ black hole from quasinormal mode. Furthermore Bekenstein's proposal holds only for non-extremal black holes.
However we might expect that the area spectrum of an extremal black hole is also equally spaced.  Indeed, in string theory we find that the entropy spectrum of the extremal black hole involves the square roots of integer quantum numbers. This looks different from the equally spaced entropy spectrum of non-extremal black holes in the semiclassical limit. However  when we consider the semiclassical limit, i.e. a large quantum number for fixed other quantum numbers,  we can find that the entropy spectrum of the extremal black hole in string theory is also equally spaced for a large quantum number.

Following the argument in this paper, we expect that for black holes with  multiple horizons  the inner horizon area as well as  the outer horizon area would be quantized and probably equally spaced in the cases of other black holes of the Einstein's gravity. If there is enough information about the transition frequency from quasinormal modes,  the quantizations of action variables will indicate how the outer and inner horizon areas are quantized.  It will be interesting to check  if  the same thing happens for the Kerr black hole and the charged black hole which have two horizons. The paper about them is in preparation \cite{sk2}. 
%

\ack{This work was supported by a grant from the Kyung Hee University in 2008.}


\end{document}